\documentclass[a4paper,12pt,dvips]{article}
\usepackage{amssymb}
\usepackage{amsmath}
\usepackage{amscd}
\usepackage{latexsym}
\usepackage{epsfig}
\usepackage{multirow}
\usepackage{color}
\usepackage[bf,footnotesize]{caption}
\usepackage{subfigure}

\usepackage[dvips,colorlinks=true,a4paper=true,backref=false, linktocpage=true,
citecolor=blue,urlcolor=blue,linkcolor=blue,pdfpagemode=UseOutlines]{hyperref}

\hypersetup{%
  bookmarksnumbered=true,
  pdftitle = {HMC algorithm with multiple time scale
    integration and mass preconditioning},
  pdfsubject = {HMC algorithm for lattice QCD },
  pdfauthor = {C.Urbach, K.Jansen, A.Shindler, U.Wenger},
  pdfkeywords = {HMC, Hybrid Monte Carlo, 
    Multiple time scale integration, mass preconditioning}
}

\graphicspath{{figures/}}

{\\{\bf ENDFIXME}\par\medskip}
{}

\renewcommand{\sectionmark}[1]%
{\markright{\thesection\ #1}}

\newcommand{\tr}{\operatorname{Tr}}

\newcommand{\Wpm}{W^{\pm}}
\newcommand{\Wpmi}[1]{W^{\pm}_{#1}}

\newcommand{\Wp}{W^{+}}
\newcommand{\Wpi}[1]{W^{+}_{#1}}
\newcommand{\Wm}{W^{-}}
\newcommand{\Wmi}[1]{W^{-}_{#1}}

\newcommand{\dtau}{\Delta\hspace{-.06cm}\tau}
\newcommand{\npf}{N_\mathrm{PF}}

\definecolor{eins}{rgb}{0,0,0}
\definecolor{zwei}{rgb}{1,0,0}
\definecolor{drei}{rgb}{0,1,0}
\definecolor{vier}{rgb}{0,0,1}
\definecolor{fuenf}{rgb}{1,1,0}
\definecolor{pink}{rgb}{1,0,1}
\definecolor{sechs}{rgb}{1,0,1}
\definecolor{sieben}{rgb}{0,1,1}
\definecolor{acht}{rgb}{0.35,0.83,0.33}
\definecolor{neun}{rgb}{0.35,0.33,0.85}

\date{}

\begin{document}

\begin{titlepage}

\title{
  {\vspace{-0cm} \normalsize
    \hfill \parbox{40mm}{DESY/05-090\\
      SFB/CPP-05-20\\June 2005}}\\[10mm]
  HMC algorithm with multiple time scale\\
  integration and mass preconditioning
}
\author{C.~Urbach$^{\, 1,\,2}$, K.~Jansen$^{\,2}$, A.~Shindler$^{\, 2}$, U.~Wenger$^{\, 2}$\\ \ \\
  {\small $^{1}$ Institut f\"{u}r Theoretische Physik, Freie Universit\"{a}t Berlin,} \\
  {\small Arnimallee 14, D-14195 Berlin, Germany} \\ \ \\
  {\small $^{2}$ John von Neumann-Institut f{\"u}r Computing NIC,} \\
  {\small Platanenallee 6, D-15738 Zeuthen, Germany} \\ 
}

\maketitle

\thispagestyle{empty}

\begin{abstract}
  \noindent{\centering
    \begin{tabular*}{\linewidth}{c}
      \hline
    \end{tabular*}}
  
  We present a variant of the HMC algorithm with mass
  preconditioning (Hasenbusch acceleration) and multiple time scale
  integration. 
  We have tested this variant for standard Wilson fermions at $\beta=5.6$ 
  and at pion masses ranging from $380$MeV to $680$MeV.
  We show that in this situation its performance is comparable to the
  recently proposed HMC variant with domain decomposition as 
  preconditioner.
  We give an update of the ``Berlin Wall'' figure, comparing the
  performance of our variant of the HMC algorithm to other published
  performance data.
  Advantages of the HMC algorithm with mass preconditioning and
  multiple time scale integration are that it is straightforward to
  implement and can be used in combination with a wide variety of
  lattice Dirac operators.
  
  \noindent{\centering
    \begin{tabular*}{\linewidth}{c}
      \hline
    \end{tabular*}}
\end{abstract}
\end{titlepage}


\section{Introduction}

Simulations of full QCD with light dynamical flavors of quarks and at
small lattice spacing are presently one of the greatest challenges in
lattice QCD. Simulations in this regime have to face the phenomenon of
critical slowing down: in addition to the ``natural'' increase of
costs due to increasing volume and increasing iteration numbers needed
in the solvers, the autocorrelation times are expected to grow
significantly when the masses are decreased.

One widely used algorithm to perform those simulations is the Hybrid
Monte Carlo (HMC) algorithm \cite{Duane:1987de}, an exact algorithm
which combines molecular dynamics evolution of the gauge fields with a
Metropolis accept/reject step to correct for discretization errors in
the numerical integration of the corresponding equations of motion.
However, in its original form the HMC algorithm is even on nowadays
computers not able to tackle simulations with light quarks on fine
lattices, for the reasons stated above.

Therefore a lot of effort has been invested to accelerate the HMC
algorithm during the last years and the corresponding list of
improvements that were found is long. It reaches, for example, from
even/odd preconditioning \cite{DeGrand:1990dk} over multiple time
scale integration \cite{Sexton:1992nu} and chronological inversion
methods \cite{Brower:1995vx} to mass preconditioning (Hasenbusch
acceleration) \cite{Hasenbusch:2001ne,Hasenbusch:2002ai}, to mention
only those that are immediately relevant for the present paper.
It is worth noting that
many of the known improvement tricks can be combined. In addition,
alternative multiboson methods \cite{Luscher:1993xx} have been
suggested, which, however, appear not to be superior to the HMC
algorithm. 

Recently in ref.~\cite{Luscher:2004rx} a HMC variant as a combination
of multiple time scale integration with domain decomposition as
preconditioner on top of even/odd preconditioning was presented and
speedup factors of about $10$ were reported when compared to state of
the art simulations using a HMC algorithm \cite{Orth:2005kq}. Even
more important, excellent scaling properties with the quark mass were
found. Thus this algorithm seems to be most promising when one wants
to simulate small quark masses on fine lattices.

In this paper we are going to present yet another variant of the HMC
algorithm similar to the one of refs.~\cite{AliKhan:2003mu,AliKhan:2003br}
comprising multiple time scale integration with mass
preconditioning on top of even/odd preconditioning. 
We test this algorithm for standard Wilson fermions at $\beta=5.6$ and
at pion masses ranging from $m_\pi=370$MeV to $m_\pi=650$MeV.
We show that in this situation the algorithm 
has similar scaling properties and performance as the method presented
in ref.~\cite{Luscher:2004rx}. From the performance data obtained with our
HMC variant we update the ``Berlin Wall'' figures of
refs.~\cite{Ukawa:2002pc,Jansen:2003nt} and find that the picture is
significantly improved.

After shortly recalling the HMC algorithm as well as the ideas of
multiple time scale integration and mass preconditioning, we will
present numerical results comparing our HMC version with the one of
refs.~\cite{Luscher:2004rx,Orth:2005kq} and draw our conclusions.



\section{HMC algorithm}

The variant of the HMC algorithm we will present here is applicable to
a wide class of lattice Dirac operators, including twisted mass
fermions, various improved versions of Wilson fermions, 
staggered fermions, and even the
overlap operator. Nevertheless, in order to discuss a concrete
example, we restrict ourselves in this paper to the Wilson-Dirac
operator with Wilson parameter $r$ set to one 
\begin{equation}
  \label{eq:WilsonDirac}
  D_\mathrm{W}[U,m_0] =
  \frac{1}{2}\sum_\mu\left\{(\nabla_\mu+\nabla_\mu^*)\gamma_\mu - a
    \nabla_\mu^* \nabla_\mu\right\} + m_0\, ,
\end{equation}
where $m_0$ is the bare mass parameter, and $\nabla_\mu$ and $\nabla_\mu^*$ the
gauge covariant lattice forward and backward difference operators,
respectively. The Wilson lattice action can be written as
\begin{equation}
  \label{eq:action}
  S[U,m_0] = S_\mathrm{G}[U] +
  a^4\sum_x\bar\psi(x)\left(D_\mathrm{W}[U,m_0]\right)\psi(x)\, ,
\end{equation}
where $S_\mathrm{G}[U]$ is the usual Wilson plaquette gauge action.
In the implementation we use the so-called hopping parameter
representation of eq.~(\ref{eq:action}), where the hopping parameter
is defined to be  
\[
\kappa = (2m_0 + 8)^{-1}\, ,
\]
and the fermion fields are rescaled according to
\[
\psi\ \to\ \frac{\sqrt{2\kappa}}{a^{3/2}}\psi\,
,\qquad\bar\psi\ \to\ \frac{\sqrt{2\kappa}}{a^{3/2}}\bar\psi\, .
\]
After integrating out the fermion fields $\psi,\bar\psi$ the
partition function of lattice QCD for $n_f=2$ mass degenerate flavors
of quarks reads
\begin{equation}
  \label{eq:partition}
  \mathcal{Z} = \int\mathcal{D}U \det(D_\mathrm{W}[U, m_0])^2
  e^{-S_\mathrm{G}[U]}\ . 
\end{equation}
Since $D_\mathrm{W}$ fulfills the property $\gamma_5D_\mathrm{W}\gamma_5 =
D_\mathrm{W}^\dagger$ we define for later purpose the hermitian
Wilson-Dirac operator
\begin{equation}
  \label{eq:Q}
  Q = \gamma_5 D_\mathrm{W}\, .
\end{equation}

In order to prepare the set up for the Hybrid Monte Carlo (HMC)
algorithm \cite{Duane:1987de}, the determinant $\det(D_\mathrm{W}[U,
m_0])$ is usually re-expressed by the use of so-called pseudo fermion
fields $\phi$ 
\begin{equation}
  \label{eq:1}
  \det(D_\mathrm{W})^2 = \det(Q)^2 \propto \int\mathcal{D}\phi^\dagger\mathcal{D}\phi
  \exp\Bigl(-S_{\mathrm{PF}}[U,\phi^\dagger,\phi]\Bigr)\, ,
\end{equation}
where $S_{\mathrm{PF}}[U,\phi^\dagger,\phi] = |Q^{-1}\phi|^2$ is the
pseudo fermion action. The pseudo fermion fields $\phi$ are formally
identical to the fermion fields $\psi$, but follow the statistic of
bosonic fields. The $\phi$ version of the HMC algorithm is based on
the Hamiltonian
\begin{equation}
  \label{hmc:hamiltonian}
  H(P, U, \phi, \phi^\dagger) = \frac{1}{2}\sum_{x,\mu}P_{x,\mu}^2 + S_\mathrm{G}[U] +
  S_{\mathrm{PF}}[U,\phi,\phi^\dagger]\, ,
\end{equation}
where we introduced traceless Hermitian momenta $P_{x,\mu}$ as conjugate
fields to the gauge fields $U_{x,\mu}$. The HMC algorithm is then
composed out of molecular dynamics evolution of the gauge fields and
momenta and a Metropolis accept/reject step, which is needed to
correct for the discretization errors of the numerical integration of
the corresponding equations of motion.





It is possible to prove that the HMC algorithm satisfies the
\emph{detailed balance} condition \cite{Duane:1987de} and hence the
configurations generated with this algorithm correctly represent the
intended ensemble.

\subsection{Molecular dynamics evolution}

In the molecular dynamics part of the HMC algorithm the gauge fields
$U$ and the momenta $P$ need to be evolved in a fictitious computer
time $t$. With respect to $t$, Hamilton's equations of motion read
\begin{equation}
  \label{eq:hmc1}
  \frac{d U}{dt} = \frac{d H}{d P}=P\, , \qquad \frac{d P}{dt} = -\frac{d
    H}{d U} = - \frac{d S}{d U}\, ,
\end{equation}
where we set $S = S_\mathrm{G} + S_{\mathrm{PF}}$ and $d/dU$, $d/dP$
formally denote the derivative with respect to group elements. Since
analytical integration of the former equations of motion is normally
not possible, these equations must in general be integrated with a
discretized integration scheme that is area preserving and reversible,
such as the leap frog algorithm. The discrete update with integration
step size $\dtau$ of the gauge field and the momenta can be defined as 
\begin{equation}
  \label{eq:hmc2}
  \begin{split}
    T_\mathrm{U}(\dtau)&:\quad U\quad\to\quad U' = \exp\left( i\dtau P\right)
      U\, ,\\
    T_\mathrm{S}(\dtau)&:\quad P\quad\to\quad P' = P - i\dtau\delta S\, ,\\
  \end{split}
\end{equation}
where $\delta S$ is an element of the Lie algebra of $\mathrm{SU}(3)$
and denotes the variation of $S$ with respect to the gauge fields. The
computation of $\delta S$ is the most expensive part in the HMC
algorithm since the inversion of the Wilson-Dirac operator is
needed. With (\ref{eq:hmc2}) one basic time evolution step of the so
called leap frog algorithm reads 
\begin{equation}
  T = T_\mathrm{S}(\dtau/2)\ T_\mathrm{U}(\dtau)\ 
  T_\mathrm{S}(\dtau/2)\, ,
\end{equation}
and a whole trajectory of length $\tau$ is achieved by
$N_\mathrm{MD}=\tau/ \dtau$ successive applications of the transformation $T$.

\subsection{Integration with multiple time scales}
\label{sec:multscales}

In order to generalize the leap frog integration scheme we assume in
the following that we can bring the Hamiltonian to the form
\begin{equation}
  \label{eq:generalH}
  H = \frac{1}{2} \sum_{x,\mu} P_{x,\mu}^2 + \sum_{i=0}^k S_i[U]\, ,
\end{equation}
with $k\geq 1$. For instance with $k=1$ $S_0$ might be identified with
the gauge action and $S_1$ with the pseudo fermion action of
eq.~(\ref{hmc:hamiltonian}). 

Given a form of the Hamiltonian (\ref{eq:generalH}) one can think of
the following situations in which it might be favorable to use a
generalized leap frog integration scheme where the different parts
$S_i$ are integrated with different step sizes, as proposed in
ref.~\cite{Sexton:1992nu}. 

Clearly, in order to keep the discretization errors in a leap frog
like algorithm small, the time steps have to be small if the driving
forces are large. Hence multiple time scale integration is a valuable
tool if the forces originating from the single parts in the
Hamiltonian (\ref{eq:generalH}) differ significantly in their absolute
values. Then the different parts in the Hamiltonian might be integrated
on time scales inverse proportionally deduced from the corresponding forces.

Another situation in which multiple time scales might be useful exists
when one part in the Hamiltonian (\ref{eq:generalH}) is significantly
cheaper to update than the others. In this case the cheap part might
be integrated without too much performance loss on a smaller time
scale, which reduces the discretization errors coming from that
part.

The leap frog integration scheme can be generalized to multiple time
scales as has been proposed in ref.~\cite{Sexton:1992nu} without loss of
reversibility and the area preserving property. The scheme with only
one time scale can be recursively extended by starting with the
definition
\begin{equation}
  T_0 = T_{\mathrm{S}_0}(\dtau_0/2)\ T_\mathrm{U}(\dtau_0)\ 
  T_{\mathrm{S}_0}(\dtau_0/2)\, ,
\end{equation}
with $T_\mathrm{U}$ defined as in eq.~(\ref{eq:hmc2}) and where
$T_{\mathrm{S}_i}(\dtau)$ is given by
\begin{equation}
  \label{eq:basicT}
  T_{\mathrm{S}_i}(\dtau)\quad : \quad P\quad\to\quad P - i \dtau\delta
  S_i[U]\, .
\end{equation}
As $\dtau_0$ will be the smallest time scale, we can recursively
define the basic update steps $T_i$, with time scales $\dtau_i$
as
\begin{equation}
  \label{eq:lfrecursive}
  T_i = T_{\mathrm{S}_i}(\dtau_i/2)\
  [T_{i-1}]^{N_{i-1}}\ T_{\mathrm{S}_i}(\dtau_i/2)\, ,
\end{equation}
with integers $N_i$ and $0<i\leq k$. One full trajectory $\tau$ is then composed
by $[T_k]^{N_k}$. The different time scales $\dtau_i$ in
eq.~(\ref{eq:lfrecursive}) must be chosen such that the total number
of steps on the $i$-th time scale $N_{\mathrm{MD}_i}$ times $\dtau_i$
is equal to the trajectory length $\tau$ for all $0\leq i\leq k$:
$N_{\mathrm{MD}_i}\dtau_i=\tau$. This is obviously achieved by setting
\begin{equation}
  \label{eq:timescales}
  \dtau_i = \frac{\tau}{N_k\cdot N_{k-1}\cdot...\cdot N_i} =
  \frac{\tau}{N_{\mathrm{MD}_i}}\, ,\qquad 0\leq i\leq k\, ,
\end{equation}
where $N_{\mathrm{MD}_i}=N_k\cdot N_{k-1}\cdot...\cdot N_i$.

In ref.~\cite{Sexton:1992nu} also a partially improved integration scheme with multiple
time scales was introduced, which reduces the size of the discretization errors.
Again, we assume a Hamiltonian of the form (\ref{eq:generalH}) with 
now $k=1$. By defining similar to $T_0$ 
\begin{equation}
  T_{\mathrm{SW}_0}\ =\ T_{\mathrm{S}_0}(\dtau_0/6)\
  T_\mathrm{U}(\dtau_0/2)\ T_{\mathrm{S}_0}(2\dtau_0/3)\
  T_\mathrm{U}(\dtau_0/2)\ T_{\mathrm{S}_0}(\dtau_0/6)\, ,
\end{equation}
the basic update step of the improved scheme -- usually referred to as
the Sexton-Weingarten (SW) integration scheme -- reads
\begin{equation}
  \begin{split}
    T_{\mathrm{SW}_1} = &T_{\mathrm{S}_1}(\dtau_1/6)\\
    &[T_{\mathrm{SW}_0}]^{N_0}\ T_{\mathrm{S}_1}(2 \dtau_1/3)\\
    &[T_{\mathrm{SW}_0}]^{N_0}\ T_{\mathrm{S}_1}(\dtau_1/6)\, ,
  \end{split}
\end{equation}
where $\dtau_0 = \dtau_1/(2N_0)$. This integration scheme not only
reduces the size of the discretization errors, 
but also sets for $S_0$  a different time scale than for $S_1$.
Hence, it is one special example for an integration scheme with
multiple time scales and can easily be extended to more than two time
scales by recursively defining ($0<i\leq k$):
\begin{equation}
  \label{eq:swrecursive}
  \begin{split}
    T_{\mathrm{SW}_i} = &T_{S_i}(\dtau_i/6)\ \\ 
    &[T_{\mathrm{SW}_{i-1}}]^{N_{i-1}}\
    T_{S_i}(2 \dtau_i/3)\\ 
    &[T_{\mathrm{SW}_{i-1}}]^{N_{i-1}}\
    T_{S_i}(\dtau_i/6)\, . 
  \end{split}
\end{equation}
The different time scales for the SW integration scheme are defined by
\begin{equation}
  \label{hmc:swtimescales}
  \dtau_i = \frac{\tau}{(2N_k)\cdot (2N_{k-1})\cdot...\cdot (2N_i)} =
  \frac{\tau}{N_{\mathrm{MD}_i}}\, ,\qquad i\leq k\, .
\end{equation}
Note that the SW partially improved integration scheme 
makes use of the fact that the computation of the variation of
the gauge action is cheap compared to the variation of the
pseudo fermion action and in addition the time scales are chosen in
order to cancel certain terms in the discretization error exactly.


\setcounter{equation}{0}

\section{Mass Preconditioning}

The arguments presented in this section are made for simplicity only
for the not even/odd preconditioned Wilson-Dirac operator. The
generalization to the even/odd preconditioned case is simple and can
be found in ref.~\cite{Hasenbusch:2001ne} and the appendix of
ref.~\cite{Farchioni:2004us}.

Mass preconditioning \cite{Hasenbusch:2001ne} -- also known as
Hasenbusch acceleration -- relies on the observation that one can
rewrite the fermion determinant as follows
\begin{equation}
  \label{eq:massprecon}
  \begin{split}
    \det(Q^2) &= \det(\Wp\Wm) \frac{\det(Q^2)}{\det(\Wp\Wm)} \\
    &=
    \int\mathcal{D}\phi_1^\dagger\mathcal{D}\phi_1\
    \mathcal{D}\phi_2^\dagger\mathcal{D}\phi_2 
    \ e^{-\phi_1^\dagger\frac{1}{\Wp\Wm}\phi_1 -
      \phi_2^\dagger \Wp\frac{1}{Q^2}\Wm\phi_2}\\ 
    &=
    \int\mathcal{D}\phi_1^\dagger\mathcal{D}\phi_1\
    \mathcal{D}\phi_2^\dagger\mathcal{D}\phi_2 
    \ e^{-S_{\textrm{PF}_1} -S_{\textrm{PF}_2}}\, . 
  \end{split}
\end{equation}
The preconditioning operators $\Wpm$ can in principle be freely
chosen, but in order to let the preconditioning work $\Wp\Wm$ should
be a reasonable approximation of $Q^2$, which is, however, cheaper to
simulate. Moreover, to allow for Monte Carlo simulations,
$\det(\Wp\Wm)$ must be positive. The generalized Hamiltonian
(\ref{hmc:hamiltonian}) corresponding to eq.~(\ref{eq:massprecon})
reads
\begin{equation}
  \label{eq:Hmass}
    H = \frac{1}{2}\sum_{x,\mu}P_{x,\mu}^2 + S_\textrm{G}[U] 
    + S_{\textrm{PF}_1}[U,\phi_1,\phi_1^\dagger] +
    S_{\textrm{PF}_2}[U,\phi_2,\phi_2^\dagger]\, , 
\end{equation}
and it can of course be extended to more than one additional field.

Note that a similar approach was presented in ref.~\cite{Clark:2004cq},
in which the introduction of $n$ pseudo fermion fields was coupled with
the $n$-th root of the fermionic kernel.

One particular choice for $\Wpm$ is given by
\begin{equation}
  \label{eq:Wpm}
  \Wpm = Q \pm i\mu\, ,
\end{equation}
with mass parameter $\mu$ refered to as a twisted mass parameter. In
fact the operator
\begin{equation}
  \label{precon:1}
  \begin{pmatrix}
    \Wp & \\
    & \Wm \\
  \end{pmatrix}
  = 
  \begin{pmatrix}
    Q & \\
    & Q \\
  \end{pmatrix}
+ i \mu\tau_3
\end{equation}
is the two flavor twisted mass operator with $\tau_3$ the third Pauli
matrix acting in flavor space. One important property of this choice
is that $\Wp\Wm = Q^2 + \mu^2$. Note that $(\Wp)^\dagger = \Wm$ and we
remark that in general also $Q$ can be a twisted mass operator.

In ref.~\cite{Hasenbusch:2002ai,DellaMorte:2003jj} it was argued that
the optimal choice for 
$\mu$ is given by $\mu^2 = \sqrt{\lambda_\mathrm{max}\lambda_\mathrm{min}}$. Here 
$\lambda_\mathrm{max}$ ($\lambda_\mathrm{min}$) is the maximal (minimal)
eigenvalue of $Q^2$. The reason for the above quoted choice is as
follows: the condition number of $Q^2+\mu^2$ is approximately
$\lambda_\mathrm{max}/ \mu^2$ and the one of $Q^2/(Q^2+\mu^2)$ approximately
$\mu^2/ \lambda_\mathrm{min}$. With $\mu^2 =
\sqrt{\lambda_\mathrm{max}\lambda_\mathrm{min}}$ these two condition numbers are
equal to $\sqrt{\lambda_\mathrm{max}/\lambda_\mathrm{min}}$, both of them being 
much smaller than the condition number of $Q^2$ which is
$\lambda_\mathrm{max}/ \lambda_\mathrm{min}$. 

Since the force contribution in the molecular dynamics evolution is
supposed to be proportional to some power of the condition number, the
force contribution from the pseudo fermion part in the action is
reduced and therefore the step size $\dtau$ can be increased, in
practice by about a factor of $2$ \cite{Hasenbusch:2001ne,Hasenbusch:2002ai}.
Therefore $Q^2$ must be inverted only about half as often as before
and if the inversion of $\Wp\Wm$, which is needed to compute $\delta
S_{\mathrm{PF}_1}$, is cheap compared to the one of $Q^2$  the
simulation speeds up by about a factor of two
\cite{Hasenbusch:2001ne,Hasenbusch:2002ai}.

One might wonder why the reduction of the condition number from $K$ to
$\sqrt{K}$ gives rise to only a speedup factor of about $2$. One 
reason for this is that one cannot make use of the reduced condition
number of $Q^2/(Q^2+\mu^2)$ in the inversion of this operator,
because in the actual simulation still the badly conditioned operator
$Q^2$ must be inverted to compute the variation of
$S_{\textrm{PF}_2}=\phi_2^\dagger\frac{\Wp\Wm}{Q^2}\phi_2$. 

\subsection{Mass preconditioning and multiple time scale integration}
\label{sec:massprecon}

In the last subsection we have seen that mass preconditioning is
indeed an effective tool to change the condition numbers of the single
operators appearing in the factorization (\ref{eq:massprecon})
compared to the original operator. But, this
reduction of the condition numbers only influences the forces -- which
are proportional to some power of the condition numbers of the
corresponding operators -- and \emph{not} the number of iterations to
invert the physical operator $Q^2$.

Therefore it might be advantageous to change the point of view: instead
of tuning the condition numbers in a way {\`a} la
refs.~\cite{Hasenbusch:2001ne,Hasenbusch:2002ai} we will exploit the
possibility of arranging the forces by the help of mass
preconditioning with the aim to arrange for a situation in which a
multiple time scale integration scheme is favorable, as explained at the 
beginning of subsection \ref{sec:multscales}.

The procedure can be summarized as follows: use mass preconditioning
to split the Hamiltonian in different parts. The forces
of the single parts should be adjusted by tuning the preconditioning
mass parameter $\mu$ such that the more expensive the computation of
$\delta S_{\mathrm{PF}_i}$ is, the less it contributes to the total
force. This is possible because the variation of $(Q^2+\mu^2)/Q^2$ is,
for $|\mu|<1$, reduced by a factor $\mu^2$ compared to the variation
of $1/Q^2$. In addition, $\Wp\Wm=Q^2+\mu^2$ is significantly cheaper
to invert than $Q^2$. Then integrate the different parts on time
scales chosen according to the magnitude of their force contribution. 

The idea presented in this paper is very similar to the idea of
separating infrared and ultraviolet modes as proposed in
ref.~\cite{Peardon:2002wb}. This idea was applied to mass
preconditioning by using only two time scales in
refs.~\cite{AliKhan:2003mu,AliKhan:2003br} in the context of clover
improved Wilson fermions. However, a comparison of our results
presented in the next section to the ones of
refs.~\cite{AliKhan:2003mu,AliKhan:2003br} is not possible, because
volume, lattice spacing and masses are different. 


\setcounter{equation}{0}

\section{Numerical results}

\subsection{Simulation points}

In order to test the HMC variant introduced in the last sections, we
decided to compare it with the algorithm proposed and tested in
ref.~\cite{Luscher:2004rx}. To this end we performed simulations with
the same parameters as have been used in ref.~\cite{Luscher:2004rx}:
Wilson-Dirac operator with plaquette gauge action at $\beta=5.6$ on
$24^3\times32$ lattices. We have three simulation points $A$, $B$ and
$C$ with values of the hopping parameter $\kappa=0.1575$,
$\kappa=0.1580$ and $\kappa=0.15825$, respectively. The trajectory
length was set to $\tau=0.5$. The details of the physical parameters
corresponding to the different simulation points can be found in table
\ref{tab:physparameter}. Additionally, this choice of simulation
points allows at the two parameter sets $A$ and $B$ a comparison to
results published in ref.~\cite{Orth:2005kq}, where a HMC algorithm with a
plain leap frog integration scheme was used.

\begin{table}[t]
  \centering
  \begin{tabular*}{.8\textwidth}{@{\extracolsep{\fill}}lccccc}
    \hline\hline
    $\Bigl.\Bigr.$ & $\kappa$ & $m_q/ \mathrm{MeV}$ & $m_\mathrm{PS}/ \mathrm{MeV}$ &
    $m_\mathrm{V}/ \mathrm{MeV}$ & $r_0/a$ \\ 
    \hline\hline
    $\Bigl.\Bigr.A$ & $0.1575$  & $66(3)$ & $665(17)$ & $947(20)$ & $6.04(10)$ \\

    $\Bigl.\Bigr.B$ & $0.1580$  & $34(1)$ & $485(13)$ & $836(24)$ & $6.18(07)$ \\

    $\Bigl.\Bigr.C$ & $0.15825$ & $22(1)$ & $380(17)$ & $839(33)$ & $6.40(15)$ \\
    \hline\hline
  \end{tabular*}
  \caption[Physical parameters of the three simulation
  points.]
    {The
    (unrenormalized) quark mass $m_q$, the pseudo scalar mass 
    $m_\mathrm{PS}$ and the vector mass $m_\mathrm{V}$ are given in
    in physical units at
    the three simulation points $A$, $B$
    and $C$. We use Wilson fermions at $\beta=5.6$ on $24^3\times32$ lattices. 
    The scale was set by the use of
    $r_0=0.5\,\mathrm{fm}$ and we give the value of $r_0/a$ at each
    simulation point. The values of all the quantities
    agree within the errors with the numbers quoted in
    refs.~\cite{Luscher:2004rx,Bali:2000vr,Orth:2005kq}, apart from
    the value for $r_0/a$ at simulation point $B$, which disagrees by
    two sigmas to the value quoted in ref.~\cite{Bali:2000vr}. This is
    presumably due to the different methods to measure this quantity.
    For the measurements we used at each simulation point $100$
    thermalized configurations separated by $5$ trajectories.
  }
  \label{tab:physparameter}
\end{table}

\subsection{Details of the implementation}

We have implemented a HMC algorithm for two flavors of mass degenerate quarks
with even/odd preconditioning and mass preconditioning with up to
three pseudo fermion fields. The boundary conditions are
periodic in all directions apart from anti-periodic ones for the
fermion fields in time direction. For details of the implementation
see the appendix of ref.~\cite{Farchioni:2004us}. For the gauge action
the usual Wilson  plaquette gauge action is used. The implementation
is written in C and uses double precision throughout.

For the mass preconditioning we use 
\begin{equation}
  \Wpmi{j} = \gamma_5(D_\mathrm{W}[U,m_0] \pm i\mu_j\gamma_5)\, ,
\end{equation}
with $j=1,2$ for the factorization in eq.~(\ref{eq:massprecon}), where
the $\mu_j$ are the additional (unphysical) twisted mass
parameters. Therefore, the pseudo fermion actions $S_{\textrm{PF}_j}$
are given by
\begin{equation}
    S_{\textrm{PF}_j}[U] = 
    \begin{cases}
      \biggl. \phi_1^\dagger\left( \frac{1}{\Wpi{1}\Wmi{1}} \right)\phi_1 &
      j=1\,\biggr. ,\\  
      \biggl. \phi_j^\dagger\left( \frac{\Wpi{j-1}\Wmi{j-1}}{Q^2}
      \right)\phi_j & j=\npf\biggr. \, ,\\ 
      \biggl. \phi_j^\dagger\left(
        \frac{\Wpi{j-1}\Wmi{j-1}}{\Wpi{j}\Wmi{j}}
      \right)\phi_j\biggr. & \text{otherwise}\, ,\\
    \end{cases}
\end{equation}
where we always chose $0<\mu_1<\mu_2$ and $\npf$ denotes the
actually used number of pseudo fermion fields. 

We have implemented the leap frog (LF) and the Sexton-Weingarten (SW)
integration schemes with multiple time scales each as described by
eq.~(\ref{eq:lfrecursive}) and eq.~(\ref{eq:swrecursive}),
respectively, where $k$ in both equations has to be identified
with $\npf$. 

The time scales are defined as in eq.~(\ref{eq:timescales}) for the LF
integration scheme and as in  eq.~(\ref{hmc:swtimescales}) for the SW
scheme, with $N_0$ corresponding to the gauge action and $N_j$ to
$S_{\mathrm{PF}_j}$ for $\npf\geq j>0$. Note that for the LF integration
scheme for one trajectory there are $N_{\npf}\cdot \ldots\cdot N_j+1$ inversions of
the corresponding operator needed, while for the SW integration scheme
there are $2N_{\npf}\cdot\ldots\cdot 2N_j+1$ inversions needed. 

For the inversions we used the CG and the BiCGstab iterative solvers.
We have tested the performance of several iterative solvers for the
even/odd preconditioned twisted mass operator
\cite{xlf:2005a} with the result, that the CG
iterative solver is the best choice in presence of a twisted
mass. Thus we used for all inversions of mass preconditioning operators
exclusively the CG iterative solver.

For the pure Wilson-Dirac operator $D_\mathrm{W}$ the BiCGstab
iterative solver is known to perform best \cite{Frommer:1994vn}. In
case of dynamical simulations, however, usually the squared hermitian
operator needs to be inverted and in this case the CG is comparable to
the BiCGstab. Only in the acceptance step, where 
$\gamma_5D_\mathrm{W}$ (or rather the even/odd preconditioned version of
it) needs to be inverted to a high precision, 
the usage of the CG would be wasteful. For
this paper we used the BiCGstab iterative solver for all inversions
of either the pure Wilson-Dirac operator itself or $(\gamma_5D_\mathrm{W})^2$.

The accuracy in the inversions was set during the computation of $\delta
S_{\mathrm{PF}_j}$ to $\epsilon_j$, which means that the inversions were
stopped when the approximate solution $\psi_j$ of 
$A_j  \psi_j=\phi_j$ fulfills
\[
\frac{\|\phi_j - A_j \psi_j \|}{\|\phi_j\|} \leq \epsilon_j \, ,
\]
where $A_j$ denotes the operator corresponding to
$S_{\mathrm{PF}_j}$. During the inversions needed for the acceptance
step the accuracy was set to $\tilde\epsilon = 10^{-10}$ for all pseudo
fermion actions. The inversions in the acceptance step must be rather
precise in order not to introduce systematic errors in the simulation,
while for the force computation the precision can be relaxed as long
as the reversibility violations are not too large. The values of
$\epsilon_j$ and $\tilde\epsilon$ have been set such 
that the reversibility violations, which should be under control
\cite{Jansen:1996cq,Liu:1997fs,Edwards:1996vs,urbach:2002aa}, are on
the same level as reported in ref.~\cite{Luscher:2004rx}, which means that
the differences in the Hamiltonian are of the order of $10^{-5}$. The
values for $\epsilon_j$ can be found in table \ref{tab:parameter}. 

The errors and autocorrelation times were computed with the so called
$\Gamma$-method as explained in ref.~\cite{Wolff:2003sm} (see also
ref.~\cite{Madras:1988ei}), i.e.
\begin{equation}
  \label{results:tauint}
  \tau_\mathrm{int} = \frac{1}{2} +
  \sum_{t=1}^\infty \frac{\Gamma(t)}{\Gamma(0)}\, ,
\end{equation}
with the autocorrelation function $\Gamma(t)$.

\begin{table}[t]
  \centering
  \begin{tabular*}{.99\textwidth}{@{\extracolsep{\fill}}lccccccc}
    \hline\hline
    $\Bigl.\Bigr.$ & Int. & $\npf$ & $N_\mathrm{therm}$ &
    $N_{\{\mathrm{0,1,2,3}\}}$ & $\epsilon_1, \epsilon_2, \epsilon_3$ & $\mu_1,\mu_2$ &
    $P_\mathrm{acc}$ \\  
    \hline\hline
    $\Bigl.\Bigr.A$ & SW & $3$ & $600$ & $3,2,1,3$ &
    $10^{-7},10^{-8},10^{-8}$ & $0.29,0.057$ & $0.86$\\ 

    $\Bigl.\Bigr.B$ & SW & $3$ & $1000$ & $3,2,1,3$ &
    $10^{-8},10^{-8},10^{-8}$ & $0.25,0.057$ & $0.81$\\

    $\Bigl.\Bigr.C$ & LF & $2$ & $1500$ & $5,6,10,$ - &
    $10^{-8},10^{-8},$ - & $0.054,$ - & $0.80$\\
    \hline\hline
  \end{tabular*}
  \caption[HMC algorithm parameters.]
  {HMC algorithm parameters for the three simulation
    points. We give the integration scheme, the number of pseudo
    fermion fields $\npf$, the number $N_\mathrm{therm}$ of trajectories
    of length $0.5$ used to thermalize the systems,
    the number $N_i$ of molecular dynamics steps for the multiple time scale
    integration scheme, the residues $\epsilon_i$ used in the solver for the
    force computation, the preconditioning mass parameter $\mu_i$ and
    the acceptance rate. We remind that $N_0$ corresponds to the gauge
    action.} 
  \label{tab:parameter}
\end{table}

\subsection{Force contributions}

The force contributions to the total force from the separate parts in
the action we label by $F_G$ for the gauge action and by $F_j$ for the
pseudo fermion action $S_{\mathrm{PF}_j}$. Since the variation of the
action with respect to the gauge fields is an element of the Lie
algebra of $\mathrm{SU}(3)$, we used $\|X\|^2 = -2 \tr{X^2}$ as the
definition of the norm of such an element.

\begin{figure}[t]
  \centering
  \subfigure[Forces for run $B$.] 
  {\label{fig:force4a}\includegraphics[width=.48\linewidth]
    {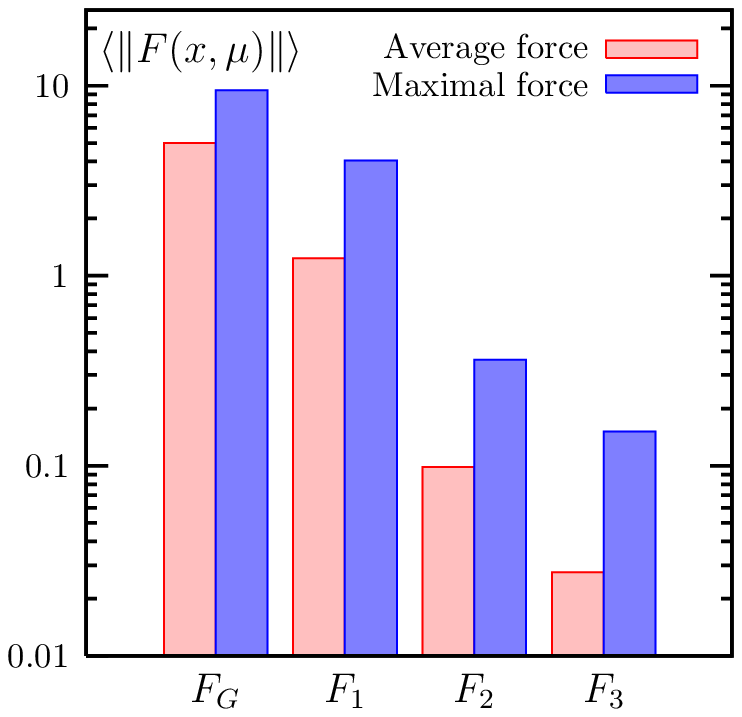}}
  \subfigure[Forces for run $C$.]
  {\label{fig:force4b}\includegraphics[width=.48\linewidth]
    {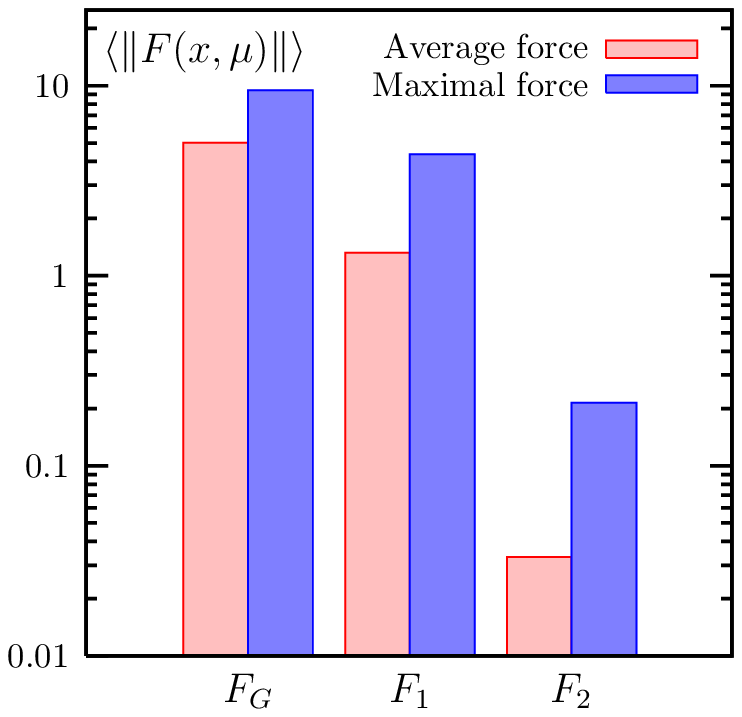}}
  \caption{Average and maximal forces for simulation points $B$ and
    $C$. The statistical errors are too small to be visible due to the
    large number of measurements.}
  \label{fig:forces}
\end{figure}

In order to better understand the influence of mass preconditioning on
the HMC algorithm we computed the average and the maximal norm of the
forces $F_G, F_1, F_2$ and $F_3$ on a given gauge field after all
corresponding gauge field updates:
\begin{equation}
  \label{results:forcedef}
  \begin{split}
    \|F\|_\mathrm{aver} &= \frac{1}{4 L^3 T}\sum_{x,\mu}\|F(x,\mu)\|\,
    ,\\
    \|F\|_\mathrm{max} &= \max_{x,\mu}\{\|F(x,\mu)\|\}\, ,\\
  \end{split}
\end{equation}
and averaged them over all measurements, which we indicate with
$\langle .\rangle$. Examples of force distributions for different runs
can be found in figure \ref{fig:forces}. These investigations lead to
the following observations generic to our simulation points: 
\begin{itemize}
\item With the choice of parameters as given in table
  \ref{tab:parameter} the single force contributions are strictly
  hierarchically ordered with\\ 
  $\|F_G\|_\mathrm{aver,max} > \|F_1\|_\mathrm{aver,max} >
  \|F_2\|_\mathrm{aver,max} > \|F_3\|_\mathrm{aver,max}$.  
\item The maximal force is up to one order of magnitude larger than
  the average force. This can only be explained by large local
  fluctuations in this quantity. These fluctuations become larger the
  smaller the mass is.
\end{itemize}
Moreover, the force ordering and sizes look very similar to the one
reported in ref.~\cite{Luscher:2004rx}. 

\begin{figure}[t]
  \centering
  \includegraphics[width=.8\linewidth]{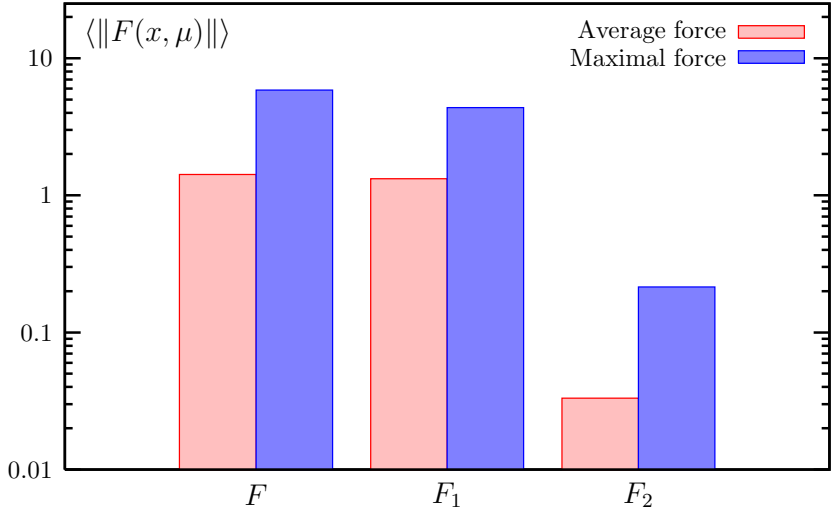}
  \caption{Comparison between the fermionic forces of run $C$ ($F_1$
    and $F_2$) and a run with $\kappa=0.15825$ without mass preconditioning
    and multiple time scales ($F$). The statistical errors are too
    small to be visible.}
  \label{fig:compC}
\end{figure}

In a next step we performed some test trajectories without mass
preconditioning in order to compare the fermionic forces with and
without mass preconditioning. For the value of $\kappa=0.15825$ (run $C$)
the result can be found in figure \ref{fig:compC}. The bars labeled
with $F$ correspond to the fermion force without mass
preconditioning. The labels $F_1$ and $F_2$ refer to the two fermionic
forces for the run $C$ with mass preconditioning. The following ratios
are of interest:
\[
\begin{split}
  &  \frac{\|F\|_\mathrm{aver}}{\|F_1\|_\mathrm{aver}} \approx 1 \,
  ,\qquad\frac{\|F\|_\mathrm{aver}}{\|F_2\|_\mathrm{aver}} \approx 42\, ,\\
  &  \frac{\|F\|_\mathrm{max}}{\|F_1\|_\mathrm{max}} \approx 1.3 \,
  ,\qquad\frac{\|F\|_\mathrm{max}}{\|F_2\|_\mathrm{max}} \approx 29\, .\\
\end{split}
\]
These ratios show that the average and
maximal norm of $F_2$ is strongly reduced compared to the average and
maximal norm of $F$. We observe that the maximal norm is slightly less
reduced than the average norm and, by varying $\mu_1$, we could confirm 
that the norm (average and maximal) of $F_2$ is roughly proportional to 
$\mu_1^2$. 

As a further observation, one sees from figure \ref{fig:compC} or
from the ratios quoted above that the norm of $F_1$ is almost
identical to the norm of $F$, which is the case for both the average
and the maximal values. 

From these investigations we think one can conclude the following: in the
first place it is possible to tune the value of $\mu_1$ (and possibly
$\mu_2$) such that the most expensive force contribution of $F_2$ (or
$F_3$) to the total force becomes small. Secondly, since in the
example above the force contributions for $F$ and $F_1$ are almost
identical -- even though the masses are very different -- we conclude
that the norm of the forces does not explain the whole 
dynamics of the HMC algorithm. For this point see also the discussion
in the forthcoming subsection. 

\subsection{Tuning the algorithm}

As mentioned already in section \ref{sec:massprecon} the tuning of
the different mass parameters and time scales could become a delicate
task. Therefore we decided to tune the parameters $\mu_1$ and possibly
$\mu_2$ such that the molecular dynamics steps number $N_{\npf}$ for
the LF or $2N_{\npf}$ for the SW integration scheme -- the number of
inversions of the original Wilson Dirac operator in the course of one
trajectory -- is about the same as the corresponding value in
ref.~\cite{Luscher:2004rx}. The values we have chosen for the mass 
parameters $\mu_i$ and the step numbers $N_i$ can be found in table
\ref{tab:parameter} and one can see by comparing to
ref.~\cite{Luscher:2004rx} that the step numbers $N_i$ (or $2N_i$) are
indeed quite similar. 

\begin{table}[t]
  \centering
  \begin{tabular*}{.8\textwidth}{@{\extracolsep{\fill}}lccccc}
    \hline\hline
    $\Bigl.\Bigr.$& $\kappa$ & $N_\mathrm{meas}$ & $\langle P\rangle$ & $\tau_\mathrm{int}(P)$\\
    \hline\hline
    $\Bigl.\Bigr.A$ & $0.1575$ & $740$ & $0.57250(3)$ & $6(2)$\\
    $\Bigl.\Bigr.B$ & $0.1580$ & $1020$ & $0.57339(3)$ & $7(2)$\\
    $\Bigl.\Bigr.C$ & $0.15825$ & $905$ & $0.57384(4)$ & $10(4)$\\
    \hline\hline
  \end{tabular*}
  \caption{For the three runs this table contains the number of
    measurements for the plaquette $N_\mathrm{meas}$, the mean plaquette
    expectation values and the corresponding autocorrelation times.}
  \label{tab:plaq}
\end{table}

The computation of the variation of $S_\mathrm{G}$ is, compared to the
variations of the other action parts, almost negligible in terms of
computer time. Therefore we set $N_0$ always large enough to ensure
that the gauge part does not influence the acceptance rate negatively
and we leave the gauge part out in the following discussion.

If one compares e.g.~for simulation point $C$ the average norm of the
fermionic forces, then one finds that it increases like $1:40$
($\|F_2\|:\|F_1\|$). The maximal norm of the forces is
accordingly strongly ordered, approximately like $1:20$. The
corresponding relations in the step numbers we had to choose (see the
values in table \ref{tab:parameter}) increase only like
$1:6$. 

Therefore we conclude that the norm of the forces can indeed 
serve as a first
criterion to tune the time scales and the preconditioning masses, by
looking for a situation in which $\dtau_i \|F_i\|_\mathrm{max}$ is a
constant independent of $i$. But, it cannot be the only criterion. 
Finally, the acceptance rate is determined by $\langle \exp(- \Delta
H) \rangle$, which depends in a more complicated way on the forces,
see e.g. ref.~\cite{Gupta:1988js}.

\begin{figure}[t]
  \centering
  \subfigure[Monte Carlo history of $\Delta P$.]
  {\includegraphics[width=.4755\linewidth]{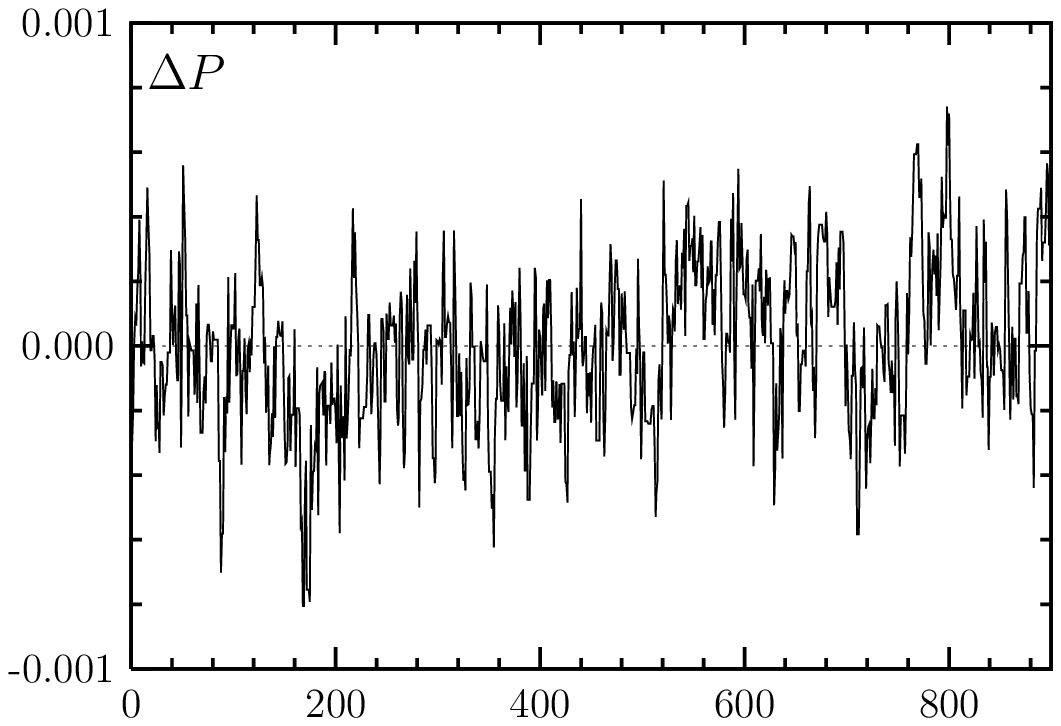}}
  \subfigure[Monte Carlo history of $\Delta H$.]
  {\includegraphics[width=.48\linewidth]{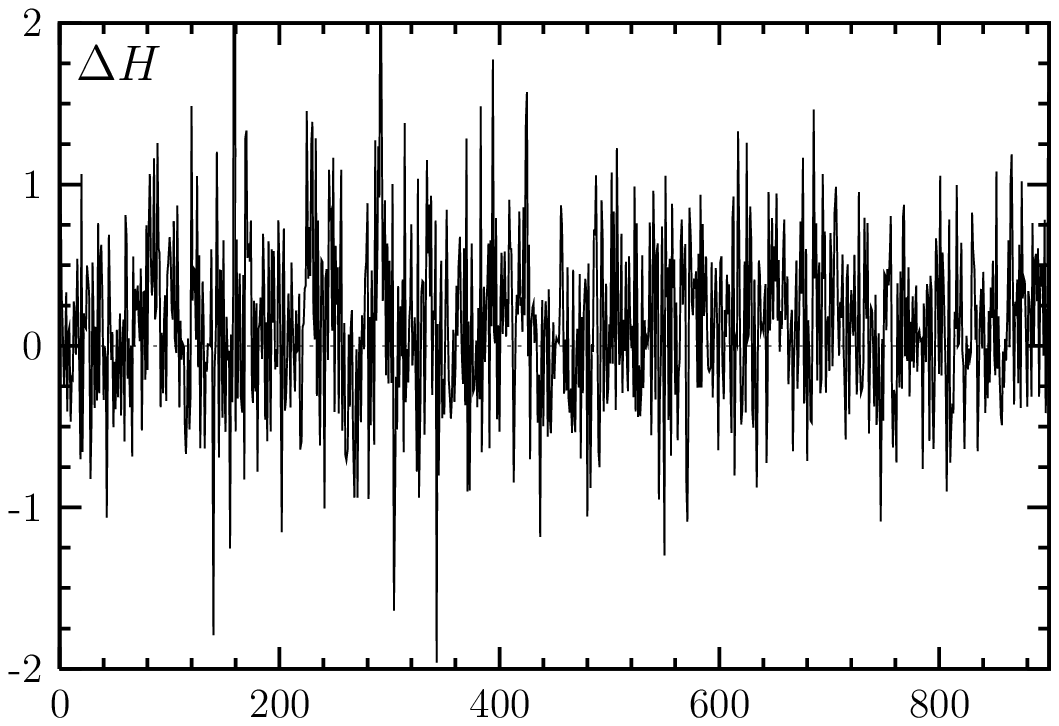}}
  \caption{Monte Carlo histories of the deviation $\Delta P$ of the average
    plaquette from its mean value and of $\Delta H$, both for simulation
    point $C$.}
  \label{fig:phistB3}
\end{figure}

It is well known that simulations with the HMC algorithm in particular for
small quark masses become often unstable if the step sizes are too
large. It is an important result that with the choice of parameters as
can be found in table \ref{tab:parameter} our simulations appear to be very
stable down to quark masses of the order of $20\,\mathrm{MeV}$. We did
encounter only few large, but not exceptional, fluctuations in $\Delta H$
during the runs. A typical history of $\Delta H$ and the average
plaquette value can be found in figure \ref{fig:phistB3} for run $C$.   
Note that even a pion mass of about $380\ \mathrm{MeV}$ might be still
to large to observe the asymptotic behavior of the algorithm.

All our runs reproduce the average plaquette expectation values quoted
in ref.~\cite{Luscher:2004rx} and, where available, in ref.~\cite{Orth:2005kq}
within the statistical errors. Our results together with the number of
measurements $N_\mathrm{meas}$ and the integrated autocorrelation time
can be found in table \ref{tab:plaq}. We also measured the values of
the pseudo scalar, the vector and the current quark mass and our
numbers agree within errors with the values quoted in
refs.~\cite{Luscher:2004rx,Orth:2005kq}. These measurements were done
on $100$ configurations separated by $5$ trajectories at each
simulation point and we computed the aforementioned quantities with
the methods explained in ref.~\cite{Farchioni:2002vn}. In order to
improve the signal we used Jacobi smearing and random sources.
Our results in physical units can be found in table
\ref{tab:physparameter}. Note that the value for $m_\mathrm{V}$ at
simulation point $C$ has to be taken with some caution, because the
lattice time extend was a bit too small to be totally sure about the
plateau. 

In order to set the scale we determined the Sommer parameter $r_0/a$
\cite{Sommer:1993ce}. Our calculation used a static action with improved
signal to noise ratio\footnote{First results applying an improved static action
  in the computation of the static potential already appeared in
  \cite{Hasenfratz:2001tw,Bali:2005fu}.}
\cite{DellaMorte:2003mn,DellaMorte:2005yc}, the tree-level improved force and
potential \cite{Sommer:1993ce} and we enhanced the overlap with the ground
state of the potential using APE smeared \cite{Albanese:1987ds} spatial gauge
links. The results can be found in table \ref{tab:physparameter}.  For run $A$
and $B$ our values for $r_0/a$ agree very well within the errors with the
value quoted in ref.~\cite{Bali:2000vr,Bali:2005fu}. One should keep in mind,
however, that the values for $r_0/a$ are computed on rather low statistics.

\subsection{Algorithm performance}

Any statement about the algorithm performance has to include 
autocorrelation times. Since different observables can have in general
rather different autocorrelation times, also the algorithm performance is
observable dependent. However, in the following we will use the
plaquette integrated autocorrelation time to determine the
performance. Note that other physical quantities such as hadron masses 
show in general very different autocorrelation times.

The values we measured for the plaquette integrated autocorrelation
times can be found in table \ref{tab:plaq}. It is interesting to
observe that for runs $A$ and $B$ the values for the plaquette
integrated autocorrelation times are smaller than the one found for
the domain decomposition method. An explanation for this may be that
in the algorithm of ref.~\cite{Luscher:2004rx} a subset of all link
variables is kept fixed during the molecular dynamics evolution, while
in our HMC variant all link variables are updated.

\begin{table}[t]
  \centering
  \begin{tabular*}{.8\textwidth}{@{\extracolsep{\fill}}lcccc}
    \hline\hline
    $\bigl.\Bigr.$& $\kappa$ & $\nu$ & $\nu$ from \cite{Luscher:2004rx} & $\nu$ from
    \cite{Orth:2005kq} \\
    \hline\hline
    $\bigl.A\Bigr.$ & $0.15750$ & $0.09(3)$ & $0.69(29)$ & $1.8(8)$\\
    $\bigl.B\Bigr.$ & $0.15800$ & $0.11(3)$ & $0.50(17)$ & $5.1(5)$\\
    $\bigl.C\Bigr.$ & $0.15825$ & $0.23(9)$ & $0.28(9)$ & -\\
    \hline\hline
  \end{tabular*}
  \caption{Values of the cost figure $\nu$ compared to the corresponding
    values of refs.~\cite{Luscher:2004rx} and \cite{Orth:2005kq}, where available.}
  \label{tab:values-cost-figure}
\end{table}

Our value for $\tau_\mathrm{int}(P)$ for run $A$ is almost identical to
the corresponding one found in ref.~\cite{Orth:2005kq}. In contrast,
for simulation point $B$ our value is a factor of three smaller, which
is only partly due to the significantly smaller acceptance rate of
about $60\%$ quoted in ref.~\cite{Orth:2005kq} for this point.

A measure for the performance of the pure algorithm, implementation
and machine independent, but incorporating the autocorrelation times is
provided by the cost figure 
\begin{equation}
  \label{eq:costfigure}
  \nu = 10^{-3}(2n + 3)\tau_\mathrm{int}(P)
\end{equation}
that has been introduced in ref.~\cite{Luscher:2004rx}. $n$ in
eq.~(\ref{eq:costfigure}) stands for either $N_{\npf}$ in case a LF
integration scheme is used or $2N_{\npf}$ in case a SW integration
scheme is used. $\nu$ represents the average number of inversions of
the Wilson-Dirac operator with the physical mass in units of thousands
as needed to generate a statistically independent value of the average
plaquette. Hence, in giving values for $\nu$, we neglect the overhead
coming from the remaining parts of the Hamiltonian.

Our values for $\nu$ together with the corresponding numbers from
ref.~\cite{Luscher:2004rx} and ref.~\cite{Orth:2005kq} are given in table
\ref{tab:values-cost-figure}. Compared to ref.~\cite{Luscher:2004rx} our
values for $\nu$ are smaller for simulation points $A$ and $B$ and
comparable for run $C$. In contrast, the cost figure for the HMC
algorithm with plain leap frog integration scheme is at least a factor
$10$ larger than the values found for our HMC algorithm variant. This
gain is, of course, what we aimed for by combining multiple time scale
integration with mass preconditioning and hence confirms our
expectation. Unfortunately, due to the large statistical uncertainties
of the $\nu$ values it is not possible to give a scaling of the cost
figure with the mass. This holds for our values of $\nu$ as well as the
ones of ref.~\cite{Luscher:2004rx}.

\subsection{Simulation cost}

\begin{table}[t]
  \centering
  \renewcommand{\multirowsetup}{\centering}
  \begin{tabular*}{.9\textwidth}{@{\extracolsep{\fill}}lccccc}
    \hline\hline
    & \multicolumn{3}{c}{$\Bigl. N_\mathrm{MV}\Bigr.$} &
    \multicolumn{2}{c}{$\tau_\mathrm{int}(P)\cdot\sum N_\mathrm{MV}$}\\
    & $\Bigl. S_{\mathrm{PF}_1}\Bigr.$ &$S_{\mathrm{PF}_2}$ &
    $S_{\mathrm{PF}_3}$ & this paper & ref.~\cite{Orth:2005kq}\\  
    \hline\hline
    $\Bigl.A\bigr.$ & $3800$ & $4600$ & $6600$ & $90000$ & $190750$\\

    $\Bigl.B\bigr.$ & $6000$ & $6900$ & $11900$ & $173600$ & $1280000$\\

    $\Bigl.C\bigr.$ & $31000$ & $25500$ & - & $565000$ & --\\
    \hline\hline
  \end{tabular*}
  \caption{Rounded number of matrix vector multiplications 
    needed during one trajectory of length $0.5$ for the different
    pseudo fermion actions without the usage of a chronological solver
    guess. We give also the sum of our numbers multiplied by the
    plaquette autocorrelation time and as a comparison the
    corresponding number from ref.~\cite{Orth:2005kq}, where available.}
  \label{tab:niter}
\end{table}

Although the value of $\nu$ is a sensible performance measure for the
algorithm itself, since it is independent of the machine, the actual
implementation and the solver, it cannot serve to
estimate the actual computer resources (costs) needed to generate 
one independent configuration. Assuming that the dominant
contribution to the total cost stems from the matrix vector (MV)
multiplications, we give in table \ref{tab:niter} the average
number of MV multiplications $N_\mathrm{MV}$ needed for the different
pseudo fermion actions to evolve the system for one trajectory of
length $\tau=0.5$. In addition we give the sum of these MV
multiplications multiplied with the plaquette autocorrelation time
together with the corresponding number from ref.~\cite{Orth:2005kq}. 

In order to compare to the numbers of ref.~\cite{Orth:2005kq} we
remark that the lattice time extent is $T=40$ in ref.~\cite{Orth:2005kq} 
compared to $T=32$ in our case, but we do not expect a large influence
on the MV multiplications coming from this small difference. Large
influence on the MV multiplications, however, we expect from ll-SSOR
preconditioning \cite{Fischer:1996th} that was used in
ref.~\cite{Orth:2005kq} in combination with a chronological solver
guess (CSG) \cite{Brower:1995vx}. 

Initially, when one compares the values of the cost figure for our HMC
algorithm with the one of the plain leap frog algorithm as
used in ref.~\cite{Orth:2005kq}, one might expect that the number of
MV multiplications shows a similar behavior as a function of the quark
mass. However, inspecting table \ref{tab:niter}, we see that in terms
of MV multiplications at simulation point $A$ the HMC algorithm of
ref.~\cite{Orth:2005kq} is only a factor of $2$ slower than the
variant presented in this paper, while the values of $\nu$ are by a
factor of about $20$ different. The reason for this is two-fold: On
the one hand ll-SSOR preconditioning together with a CSG method is
expected to perform better than only even/odd preconditioning. On the
other hand we think that the quark mass at this simulation point is
still not small enough to gain significantly from multiple time scale
integration. This illustrates that indeed the value of $\nu$ is not
immediately conclusive for the actual cost of the algorithm.

At simulation point $B$ the relative factor between the MV
multiplications needed by the two algorithms is already about $7$. And
finally, it is remarkable that for simulation point $C$ the costs with
our HMC variant are still a factor of $2$ smaller than the costs for
simulation point $B$ with the algorithm used in
ref.~\cite{Orth:2005kq}, even though the masses are very different. 

From this comparison we conclude that especially in the regime of
small quark masses the HMC algorithm presented in this paper is
significantly faster than a HMC algorithm with single time scale leap
frog integration scheme. 

By looking at table \ref{tab:niter} one notices that especially for
simulation point $C$ the number of MV multiplications needed for
preconditioning is larger than the one needed for the physical
operator. This comes from the fact that with the choice of algorithm
parameters we have used the  number of molecular dynamics steps
for the mass preconditioned operator is large. This possibly indicates
potential to further impove the performance by tuning the
preconditioning masses and time scales.

We stress here again that the number of matrix vector operations 
is highly solver dependent, and therefore, every improvement to reduce
the solver iterations will decrease the cost for one trajectory. 
Two promising improvements are the following:
\begin{itemize}
\item The use of a chronological inversion method
  \cite{Brower:1995vx}:

  The idea of the chronological inversion method (or similar methods
  \cite{Brower:1994er}) is to optimize the initial guess for
  the solution used in the solver. To this end the history of
  $N_\mathrm{CSG}$ last solutions of the equation $M^2 \chi = \phi$ is saved
  and then a linear combination of the fields $\chi_i$ with coefficients
  $c_i$ is used as an initial guess for the next inversion. $M$ stands
  for the operator to be inverted and has to be replaced by the
  different ratios of operators used in this paper.

  The coefficients $c_i$ are determined by solving 
  \begin{equation}
    \label{results:chrono}
    \sum_i \chi_j^\dagger M^2 \chi_i c_i = \chi_j^\dagger \phi
  \end{equation}
  with respect to the coefficients $c_i$. This is equivalent to
  minimizing the functional that is minimized by the CG inverter itself.
  
  In ref.~\cite{Brower:1995vx} it was reported that with a
  chronological solver guess the number of MV multiplications can be
  reduced by a factor $5$ or even more. The gain is larger the smaller
  the size of the time steps is. But at the same time the
  reversibility violations increase at equal stopping criteria in the solver.

  We have implemented the CSG method and tested its potential in
  the runs for this paper. On the one hand we see a significant
  reduction of MV multiplications on the small time scales,
  while the improvement for the large time scales is small, as
  expected.

  On the other hand we observe that the reversibility violations
  increase significantly by one or two orders of magnitude in the
  Hamiltonian when the CSG is switched on and all other parameters are
  kept fixed. Therefore one has to adjust the residues in the solvers,
  which increases the number of MV multiplications again.

  In total we found not more than a $20\%$ gain in matrix vector
  operations when a CSG is used. The largest gain is seen for the
  largest value of $\kappa$ under investigation. It is expected that this
  gain increases when the value of the bare physical mass is
  further reduced, because probably the size of the time steps must be
  further decreased. 

\item A different solver than the CG iterative solver, e.g.~a
  solver using a Schwarz method as presented in ref.~\cite{Luscher:2003qa}
  can also reduce the iteration numbers significantly. The method
  introduced in ref.~\cite{Luscher:2003qa} is expected to be particularly
  useful for inverting the original fermion matrix with a small mass.
\end{itemize}
Finally, it is interesting to compare the number of matrix vector
multiplications reported in table \ref{tab:niter} with a HMC algorithm
where mass preconditioning and multiple time scale improvements are
switched off and CSG is not used. For instance for a simulation with a
Sexton-Weingarten improved integration scheme at $\kappa=0.15825$ there are
$120$ molecular dynamics steps needed to get acceptance. This
corresponds to $240$ inversions of $Q^2$, which amounts to about
$720000$ matrix vector multiplications. Compared to run $C$ this is at
least a factor $10$ more. We did only a few trajectories to get an
estimate for this number, so we cannot say anything about
autocorrelation time.

Of course it would be interesting to compare also to a HMC algorithm
with mass preconditioning but without multiple time scale
integration. This, however, needs again a tuning of the mass
parameters and would therefore be quite costly and we did not attempt
to test this situation here.

\begin{figure}[t]
  \centering
  \subfigure[Comparison to ref.~\cite{Orth:2005kq}.]
  {\label{fig:berlin-walla} \includegraphics[width=.45\linewidth]
    {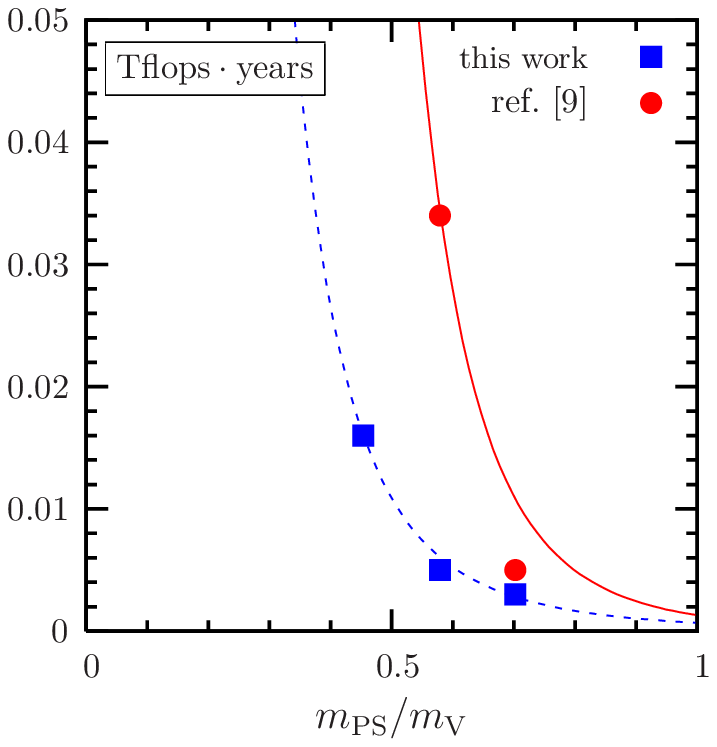}}
  \subfigure[Update of the ``Berlin Wall'' plots of
  refs.~\cite{Ukawa:2002pc,Jansen:2003nt}.]
  {\label{fig:berlin-wallb} \includegraphics[width=.443\linewidth]
    {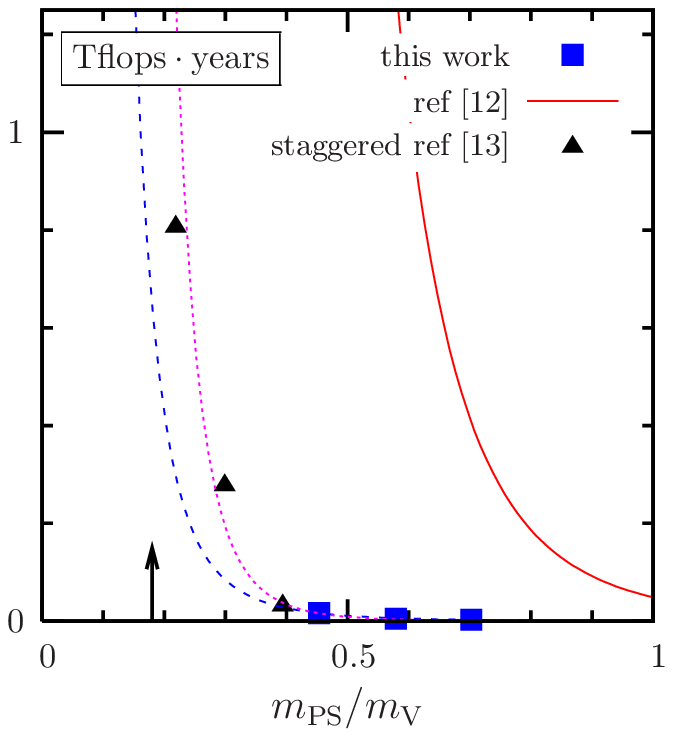}}
  \caption[Computer resources as a function of the quark mass.]
  {Computer resources needed to generate $1000$ independent
    configurations of size $24^3\times 40$ at a lattice spacing of about
    $0.08\ \mathrm{fm}$ in units of $\mathrm{Tflops}\cdot
    \mathrm{years}$ as a function of $m_\mathrm{PS}/m_\mathrm{V}$. In
    (a) we compare our results represented by squares to the results of
    ref.~\cite{Orth:2005kq} represented by circles. The lines are
    functions proportional to $(m_\mathrm{PS}/m_\mathrm{V})^{-4}$ (dashed) and
    $(m_\mathrm{PS}/m_\mathrm{V})^{-6}$ (solid) with a coefficient such
    that they cross the data points corresponding to the lightest
    pseudo scalar mass.
    In (b) we compare to the formula of eq.~\ref{results:ukawa}
    \cite{Ukawa:2002pc} (solid line) by extrapolating our data with
    $(m_\mathrm{PS}/m_\mathrm{V})^{-4}$ (dashed) and with
    $(m_\mathrm{PS}/m_\mathrm{V})^{-6}$ (dotted), respectively. The arrow
    indicates the physical pion to rho meson mass ratio. Additionally,
    we add points from staggered simulations as were used for the
    corresponding plot in ref.~\cite{Jansen:2003nt}.
    Note that all the cost data were scaled to match a lattice
    time extend of $T=40$.}
  \label{fig:berlin-wall}
\end{figure}

\subsection{Scaling with the mass}

An important property of an algorithm for lattice QCD is the scaling
of the costs with the simulated quark mass. The naive expectation is
that the number of solver iterations grows like $m_q^{-1}$ and also
the number of molecular dynamics steps is proportional to $m_q^{-1}$,
see for instance ref.~\cite{Jansen:1996xp} or
ref.~\cite{Ukawa:2002pc}. Since also the integrated autocorrelation
time is assumed to grow like $m_q^{-1}$, it is expected that the HMC
algorithm costs scale with the quark mass as $m_q^{-3}$ or
equivalently as $m_\mathrm{PS}^{-6}$.  In contrast, for our HMC 
algorithm variant we expect a much weaker scaling of  $\Delta\tau$ and also of
the number of solver iterations. Indeed, we see that the costs for our
HMC algorithm variant is consistent with a $m_q^{-2}$ or
$m_\mathrm{PS}^{-4}$ behaviour when the autocorrelation time is taken
into account.

We have translated the number of matrix vector multiplications from table
\ref{tab:niter} into costs in  units $\mathrm{Tflops}\cdot \mathrm{years}$ and
plotted the computer resources needed to generate $1000$ independent
configurations of size $24^3\times40$ at a lattice spacing of $\sim
0.08\,\mathrm{fm}$ as a function of $m_\mathrm{PS}/m_\mathrm{V}$ in figure
\ref{fig:berlin-walla} together with the results of ref.~\cite{Orth:2005kq}.
Note that we have scaled our costs like $(40/32)^{1.25}$ corresponding
to the expected volume dependence (cf.~\cite{Ukawa:2002pc}) to match the
different time extents and, moreover, we used the plaquette
autocorrelation time as an estimate for the autocorrelation time.

The solid (dashed) line is not a fit to the data, but a function
proportional to $(m_\mathrm{PS}/m_\mathrm{V})^{-4}$
($(m_\mathrm{PS}/m_\mathrm{V})^{-6}$) with a coefficient that is fixed by
the data point corresponding to the lightest pseudo scalar mass. This
functional dependencies on $(m_\mathrm{PS}/m_\mathrm{V})$ describes the
data reasonably well. However, from figure \ref{fig:berlin-walla} it
is not possible to decide on the value of the exponent in the quark
mass dependence of the costs. But, it is clear from the figure that
with multiple time scale integration and mass preconditioning the
``wall'' -- which renders simulations at some point infeasible --  is
moved towards smaller values of the quark mass. 

On a larger scale we can compare the extrapolations of our cost data
to the formula given in ref.~\cite{Ukawa:2002pc} 
\begin{equation}
  \label{results:ukawa}
  C = K \left(\frac{m_\mathrm{PS}}{m_\mathrm{V}}\right)^{-z_\pi}
  L^{z_L}\  a^{-z_a}\, , 
\end{equation}
where the constant $K$ can be found in ref.~\cite{Ukawa:2002pc} and
$z_\pi=6$, $z_L=5$ and $z_a=7$. The result of this comparison is plotted
in figure \ref{fig:berlin-wallb}, which is an update of the ``Berlin
Wall'' figure that can be found in ref.~\cite{Jansen:2003nt}. We plot
the simulation costs in units of $\mathrm{Tflops}\cdot \mathrm{years}$
versus $m_\mathrm{PS}/m_\mathrm{V}$, where we again scaled the numbers
in order to match a lattice time extend of $T=40$. The dashed and the
dotted lines are extrapolations from our data proportional to
$(m_\mathrm{PS}/m_\mathrm{V})^{-4}$ and $(m_\mathrm{PS}/m_\mathrm{V})^{-6}$,
respectively, again matching the data point corresponding to the
lightest pseudo scalar mass. The solid line corresponds to
eq.~(\ref{results:ukawa}). In addition we plot data from staggered
simulations as were used for the plot in
ref.~\cite{Jansen:2003nt}. That the corresponding points lie nearly on
top of the dotted line is accidental.

Conservatively one can conclude from figure \ref{fig:berlin-wallb}
that with the HMC algorithm described in this paper at least
simulations with $m_\mathrm{PS}/m_\mathrm{V}\approx0.3$ are feasible,
even though $L=1.93\,\mathrm{fm}$ is too small for such values of the
masses. Taking the more optimistic point of view by assuming that the
costs scale with $z_\pi=4$, even simulation with the physical
$m_\mathrm{PS}/m_\mathrm{V}$ ratio and a lattice spacing of
$0.08\,\mathrm{fm}$ become accessible, with again the caveat that
$L/a$ needs to be increased.

Independent of the value for $z_\pi$, figure \ref{fig:berlin-wallb}
reveals that the costs for simulations with staggered fermions and
with Wilson fermions in a comparable physical situation are of the
same order of magnitude, if for the simulations with Wilson fermions
an algorithm like the one presented in this work is used.
It would be interesting to see whether the techniques applied in this
paper work similarly well for staggered fermions.

We would like to point out that we did not try to tune the parameters
to their optimal values. The aim of this paper was to give a first
comparison of mass preconditioned HMC algorithm with multiple time
scale integration to existing performance data, i.e. data for a HMC
algorithm preconditioned by domain decomposition \cite{Luscher:2004rx}
and data for the HMC algorithm variant of ref.~\cite{Orth:2005kq}. We
are confident that there are still improvements possible by further
tuning of the parameters in our variant of the HMC algorithm.


\setcounter{equation}{0}

\section{Conclusion}

In this paper we have presented and tested a variant of the HMC
algorithm combining multiple time scale integration with mass
preconditioning (Hasenbusch acceleration). The aim of this paper was
to perform a first investigation of the performance properties of this
HMC algorithm by comparing it to other state of the art HMC algorithm
variants at the same situation, i.e. for bare quark masses in the
range of $20$ to $60\,\mathrm{MeV}$, a lattice spacing of about
$0.08\,\mathrm{fm}$ and a lattice size of $L\approx 2$fm 
with two flavors of mass degenerate Wilson
fermions. 

We computed at each simulation point the expectation values of
the plaquette
and of the pion, the vector and the current
quark masses finding full agreement with results in the literature. In
order to set the scale we computed the Sommer scale
\cite{Sommer:1993ce}, providing a value for $r_0/a$ also at the lowest
quark mass we simulated and which has not been available in the
literature so far. 

We have shown that the additional mass parameters introduced for mass
preconditioning can be arranged such that the force contributions from
the different parts in the Hamiltonian are strictly ordered with
respect to the absolute value of the force and that the most expensive
part has the smallest contribution to the total force.  

Using this result, it is possible to tune the time scales such that
the performance of our variant in terms of the cost figure
in eq.~(\ref{eq:costfigure}) is compatible to the one observed for the
HMC algorithm with multiple time scales and domain decomposition as 
preconditioner introduced in ref.~\cite{Luscher:2004rx} and clearly
superior to the one for the HMC algorithm with a simple leap frog
integration scheme as used in ref.~\cite{Orth:2005kq}. 

While the cost figure provides a clean algorithm
performance measure we also compare the simulation costs in units of
$\mathrm{Tflops}\cdot\mathrm{years}$ to existing data. This comparison is
summarized in an update of the ``Berlin wall'' plot of
ref.~\cite{Jansen:2003nt}, which can be found in figure 
\ref{fig:berlin-wall}. We could show that with the HMC algorithm
presented in this paper the wall is moved towards smaller values of
the quark mass and that simulations with a ratio of
$m_\mathrm{PS}/m_\mathrm{V}\approx0.3$ become feasible at a lattice spacing
of around $0.08\,\mathrm{fm}$ and $L\approx 2$fm.

The HMC variant presented here has the advantage of being applicable
to a wide variety of Dirac operators, including in principle also the
overlap operator. In addition its implementation is straightforward,
in particular in an already existing HMC code. We remark that the
paralllelization properties of our HMC variant and the one of
the algorithm presented in \cite{Luscher:2004rx} can be very different
depending on whether a fine- or a coarse-grained massively parallel
computer architecture is used.

From a stability point of view our results reveal that 
even for Wilson fermions it is very well
possible to simulate quark masses of the order of $20\ \mathrm{MeV}$
when using the algorithmic ideas presented in this
paper. We are presently simulating even smaller quark masses without
practical problems, but the statistics is not yet adequate to say
something definite.

The results presented in this paper are mostly based on empirical
observations and on simulations for only one value of the coupling
constant $\beta=5.6$. It remains to be seen how our HMC variant behaves
for larger values of $\beta$, which, as well as smaller quark masses and
theoretical considerations about the scaling properties with the quark
mass needs further investigations.

Moreover, a more systematic study of the interplay between integration
schemes, step sizes, (preconditioning and physical) masses and the
simulation costs is needed. Those investigations will hopefully also
provide a better understanding of the algorithm itself and its dynamics. 

Finally, we think that there are further improvements possible by the
usage of a Polynomial HMC (PHMC) algorithm
\cite{deForcrand:1996ck,Frezzotti:1997ym,Frezzotti:1998eu,Frezzotti:1998yp}. 
With such an algorithm one
could treat the lowest eigenvalues of the Dirac operator exactly
and/or by reweighting. In this set-up the large fluctuations in the force
might be significantly reduced, if the lowest eigenvalues are
responsible for those. Then it might be possible to further reduce the
number of inversions of the badly conditioned physical operator needed
to evolve the system.

\section*{Acknowledgements}

We thank M.~L{\"u}scher, I.~Montvay and I.~Wetzorke for helpful comments and
discussions, the qq+q collaboration and in particular F.~Farchioni, I.~Montvay
and E.E.~Scholz for providing us their analysis program for the masses.  We
thank C.~Destri and R.~Frezzotti for giving us access to a PC cluster in
Milano, where parts of the computations for this paper have been performed. We
also thank the computer-centers at HLRN and at DESY Zeuthen for granting the
necessary computer-resources, and M.~Hasenbusch for leaving us his Wilson HMC
code as a starting point. This work was supported by the DFG
Sonderforschungsbereich/Transregio SFB/TR9-03.



\bibliographystyle{h-physrev4}
\bibliography{bibliography}

\begin{thebibliography}{10}

\bibitem{Duane:1987de}
S.~Duane, A.~D. Kennedy, B.~J. Pendleton and D.~Roweth,
\newblock Phys. Lett. {\bf B195}, 216 (1987).

\bibitem{DeGrand:1990dk}
T.~A. DeGrand and P.~Rossi,
\newblock Comput. Phys. Commun. {\bf 60}, 211 (1990).

\bibitem{Sexton:1992nu}
J.~C. Sexton and D.~H. Weingarten,
\newblock Nucl. Phys. {\bf B380}, 665 (1992).

\bibitem{Brower:1995vx}
R.~C. Brower, T.~Ivanenko, A.~R. Levi and K.~N. Orginos,
\newblock Nucl. Phys. {\bf B484}, 353 (1997), [hep-lat/9509012].

\bibitem{Hasenbusch:2001ne}
M.~Hasenbusch,
\newblock Phys. Lett. {\bf B519}, 177 (2001), [hep-lat/0107019].

\bibitem{Hasenbusch:2002ai}
M.~Hasenbusch and K.~Jansen,
\newblock Nucl. Phys. {\bf B659}, 299 (2003), [hep-lat/0211042].

\bibitem{Luscher:1993xx}
M.~L{\"u}scher,
\newblock Nucl. Phys. {\bf B418}, 637 (1994), [hep-lat/9311007].

\bibitem{Luscher:2004rx}
M.~L{\"u}scher,
\newblock Comput. Phys. Commun. {\bf 165}, 199 (2005), [hep-lat/0409106].

\bibitem{Orth:2005kq}
B.~Orth, T.~Lippert and K.~Schilling,
\newblock Phys. Rev. D {\bf 72} (2005) 014503, [hep-lat/0503016].

\bibitem{AliKhan:2003mu}
A.~Ali~Khan {\em et~al.},
\newblock Nucl. Phys. Proc. Suppl. {\bf 129}, 853 (2004), [hep-lat/0309078].

\bibitem{AliKhan:2003br}
QCDSF, A.~Ali~Khan {\em et~al.},
\newblock Phys. Lett. {\bf B564}, 235 (2003), [hep-lat/0303026].

\bibitem{Ukawa:2002pc}
CP-PACS and JL{QCD}, A.~Ukawa,
\newblock Nucl. Phys. Proc. Suppl. {\bf 106}, 195 (2002).

\bibitem{Jansen:2003nt}
K.~Jansen,
\newblock Nucl. Phys. Proc. Suppl. {\bf 129}, 3 (2004), [hep-lat/0311039].

\bibitem{Farchioni:2004us}
F.~Farchioni {\em et~al.},
\newblock Eur. Phys. J. {\bf C39}, 421 (2005), [hep-lat/0406039].

\bibitem{Clark:2004cq}
M.~A. Clark and A.~D. Kennedy,
\newblock hep-lat/0409134.

\bibitem{DellaMorte:2003jj}
ALPHA, M.~Della~Morte {\em et~al.},
\newblock Comput. Phys. Commun. {\bf 156}, 62 (2003), [hep-lat/0307008].

\bibitem{Peardon:2002wb}
TrinLat, M.~J. Peardon and J.~Sexton,
\newblock Nucl. Phys. Proc. Suppl. {\bf 119}, 985 (2003), [hep-lat/0209037].

\bibitem{Bali:2000vr}
TXL, G.~S. Bali {\em et~al.},
\newblock Phys. Rev. {\bf D62}, 054503 (2000), [hep-lat/0003012].

\bibitem{xlf:2005a}
\xlf, T.~Chiarappa {\em et~al.},
\newblock in preparation  (2005).

\bibitem{Frommer:1994vn}
A.~Frommer, V.~Hannemann, B.~Nockel, T.~Lippert and K.~Schilling,
\newblock Int. J. Mod. Phys. {\bf C5}, 1073 (1994), [hep-lat/9404013].

\bibitem{Jansen:1996cq}
K.~Jansen and C.~Liu,
\newblock Nucl. Phys. Proc. Suppl. {\bf 53}, 974 (1997), [hep-lat/9607057].

\bibitem{Liu:1997fs}
C.~Liu, A.~Jaster and K.~Jansen,
\newblock Nucl. Phys. {\bf B524}, 603 (1998), [hep-lat/9708017].

\bibitem{Edwards:1996vs}
R.~G. Edwards, I.~Horvath and A.~D. Kennedy,
\newblock Nucl. Phys. {\bf B484}, 375 (1997), [hep-lat/9606004].

\bibitem{urbach:2002aa}
C.~Urbach,
\newblock Untersuchung der {R}eversibilit{\"a}tsverletzung im {H}ybrid {M}onte
  {C}arlo {A}lgorithmus,
\newblock Diploma thesis, Freie Universit{\"a}t Berlin, Fachbereich Physik,
  2002.

\bibitem{Wolff:2003sm}
ALPHA, U.~Wolff,
\newblock Comput. Phys. Commun. {\bf 156}, 143 (2004), [hep-lat/0306017].

\bibitem{Madras:1988ei}
N.~Madras and A.~D. Sokal,
\newblock J. Statist. Phys. {\bf 50}, 109 (1988).

\bibitem{Gupta:1988js}
R.~Gupta, G.~W. Kilcup and S.~R. Sharpe,
\newblock Phys. Rev. {\bf D38}, 1278 (1988).

\bibitem{Farchioni:2002vn}
F.~Farchioni, C.~Gebert, I.~Montvay and L.~Scorzato,
\newblock Eur. Phys. J. {\bf C26}, 237 (2002), [hep-lat/0206008].

\bibitem{Sommer:1993ce}
R.~Sommer,
\newblock Nucl. Phys. {\bf B411}, 839 (1994), [hep-lat/9310022].

\bibitem{Hasenfratz:2001tw}
A.~Hasenfratz, R.~Hoffmann and F.~Knechtli,
\newblock Nucl. Phys. Proc. Suppl. {\bf 106}, 418 (2002), [hep-lat/0110168].

\bibitem{Bali:2005fu}
SESAM, G.~S. Bali, H.~Neff, T.~Duessel, T.~Lippert and K.~Schilling,
\newblock Phys. Rev. {\bf D71}, 114513 (2005), [hep-lat/0505012].

\bibitem{DellaMorte:2003mn}
ALPHA, M.~Della~Morte {\em et~al.},
\newblock Phys. Lett. {\bf B581}, 93 (2004), [hep-lat/0307021].

\bibitem{DellaMorte:2005yc}
M.~Della~Morte, A.~Shindler and R.~Sommer,
\newblock hep-lat/0506008.

\bibitem{Albanese:1987ds}
APE, M.~Albanese {\em et~al.},
\newblock Phys. Lett. {\bf B192}, 163 (1987).

\bibitem{Fischer:1996th}
S.~Fischer {\em et~al.},
\newblock Comp. Phys. Commun. {\bf 98}, 20 (1996), [hep-lat/9602019].

\bibitem{Brower:1994er}
R.~C. Brower, A.~R. Levi and K.~Orginos,
\newblock Nucl. Phys. Proc. Suppl. {\bf 42}, 855 (1995), [hep-lat/9412004].

\bibitem{Luscher:2003qa}
M.~L{\"u}scher,
\newblock Comput. Phys. Commun. {\bf 156}, 209 (2004), [hep-lat/0310048].

\bibitem{Jansen:1996xp}
K.~Jansen,
\newblock Nucl. Phys. Proc. Suppl. {\bf 53}, 127 (1997), [hep-lat/9607051].

\bibitem{deForcrand:1996ck}
P.~de~Forcrand and T.~Takaishi,
\newblock Nucl. Phys. Proc. Suppl. {\bf 53}, 968 (1997), [hep-lat/9608093].

\bibitem{Frezzotti:1997ym}
R.~Frezzotti and K.~Jansen,
\newblock Phys. Lett. {\bf B402}, 328 (1997), [hep-lat/9702016].

\bibitem{Frezzotti:1998eu}
R.~Frezzotti and K.~Jansen,
\newblock Nucl. Phys. {\bf B555}, 395 (1999), [hep-lat/9808011].

\bibitem{Frezzotti:1998yp}
R.~Frezzotti and K.~Jansen,
\newblock Nucl. Phys. {\bf B555}, 432 (1999), [hep-lat/9808038].

\end{thebibliography}
\end{document}